# Putting it into Practice

*J.-P. Burnet*
CERN, Geneva, Switzerland

**Abstract**
This paper presents the latest trends in the powering of particle accelerators. A series of solutions is proposed for responding to the challenges of high performance machines. This paper covers the domains of magnetic field uncertainty, power converter control, and energy saving. This list is not exhaustive, but it does correspond to the latest innovations in the field of powering particle accelerators.

## 1 Introduction

Particle accelerators are very challenging machines in terms of powering. Many solutions had to be found to operate these machines at the required level of performance. Thanks to the experience accumulated, synchrotron performance is well above that of the first designs of the 1950s. For example, the PS machine, built in 1957 at CERN, was designed for an intensity per pulse of $10^{10}$ protons and, after 50 years and with many upgrades, in 2013 reached $3 \times 10^{13}$ protons [1]. Power performance acts directly on the global performance of the synchrotron. The stability and reproducibility of the power converters are crucial for beam performance; this explains why particle accelerators need high-precision power converters. This paper will underline some important parameters regarding machine performance and describes the associated solutions put into place.

## 2 Magnetic field uncertainty

Magnetic field uncertainty generates many difficulties when operating a machine, especially when the machine is working with different beam energies. The magnetic field is very difficult to measure at an order of magnitude of $10^{-4}$, which unfortunately is mandatory for the control of particle accelerators. The classical way to control the magnetic field is to measure the DC current delivered by power converters. Based on these measurements, the operators can determine the beam energy. In most synchrotrons, all of the magnets (dipole, quadrupole, sextupole, and correctors) are current controlled, and the beam energy is controlled by the dipole magnet current. The stability of the magnet current and its reproducibility are very important for beam control and stability. For example, a machine can suffer from tune spread due to current ripple in the quadrupole magnets. Magnetic modelling is also a key technique to improving predictions, see Fig. 1.

### 2.1 Circuit layout

The first solution to ease the operation of synchrotrons is to connect in series all of the magnets by family (dipole, quadrupole, or sextupole). Thank to this layout, the magnetic field is expected to be identical for all of the magnets of any one family, which suppresses the uncertainty of the magnetic field, depending upon their position inside the machine. The impedance of the circuits is also more inductive, which helps to reduce the current ripple. However, in some cases this layout isn't possible due to the size of the machine (as for the LHC) or due to the rigidity introduced in the beam optics (as for light sources). In these cases the quality of current measurement and current control has to be even better, to obtain the required performance for the machine [2].

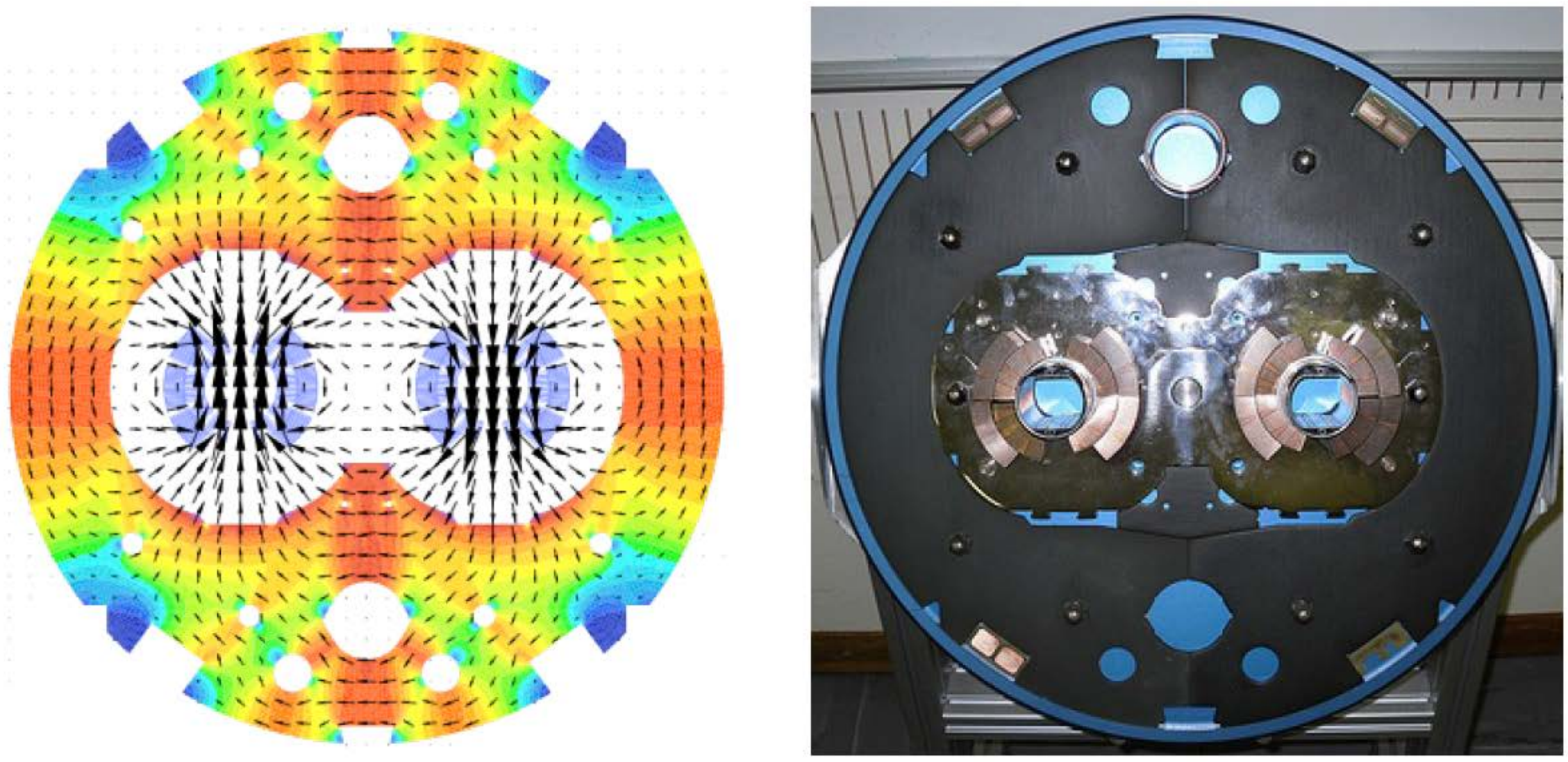

(a) (b)

**Fig. 1:** (a) Magnetic field in a dipole magnet; (b) and cross-section through a LHC dipole

## 2.2 Degauss and pre-cycle

All magnets need a proper pre-cycling to have reproducible behaviour, i.e. to provide the same field at the same level of current. Depending on the type of magnet, the type of pre-cycle stems from different physical phenomena, and reproducibility can be obtained through different pre-cycle strategies [3]. For a predicable magnetic field value, it is imperative to follow a specific path on the magnetization curves so that any cycle will arrive at a predictable and repeatable value despite the hysteresis in flux changes. Figure 2 explains how the magnets can be driven to always operate on a known magnetization curve. The idea is to use the magnetization curve from a stable point; and the only stable point is when the magnet is saturated. All degauss cycles need to go to the maximum current to start the prediction from a constant point.

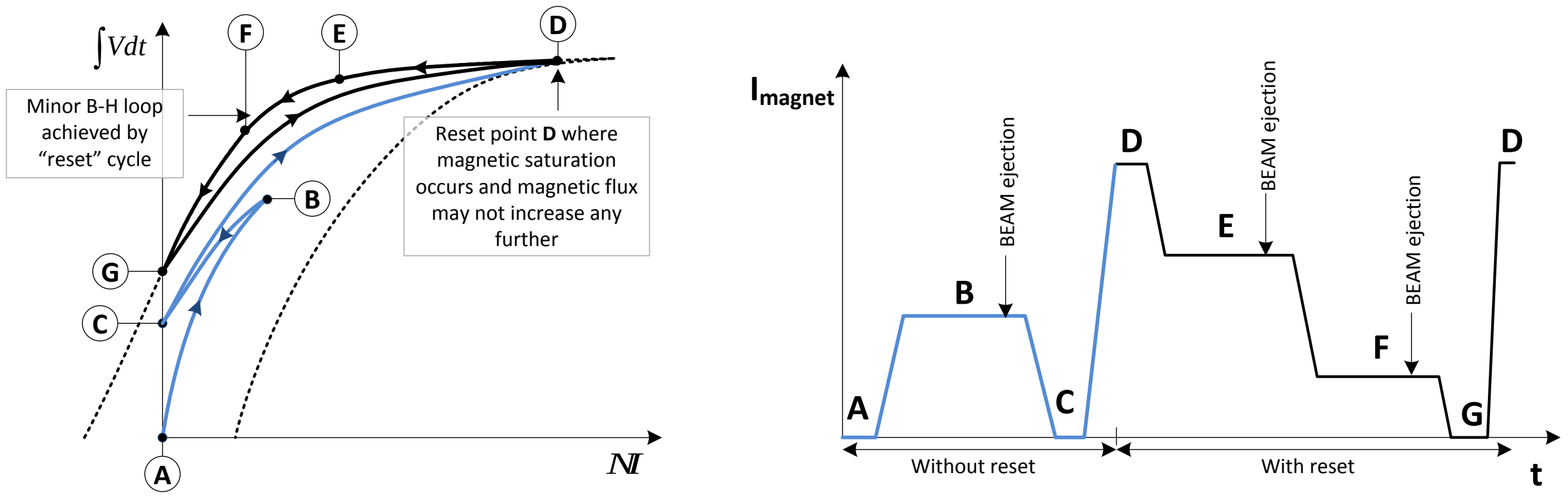

(a) (b)

**Fig. 2:** (a) The magnetic flux intensity versus magnetic flux density in a typical magnet; (b) evolving as the magnet current is varied following a typical waveform. Operation without the reset cycle is in blue. The desired ‘predictable’ operation around a minor B-H loop is in black.

Different types of pre-cycles can be executed. For magnets that always operate at the same beam energy, the classical five pre-cycles can be executed, see Fig. 3. This is the case for beam transfer lines with fixed energy and the LHC warm magnets.

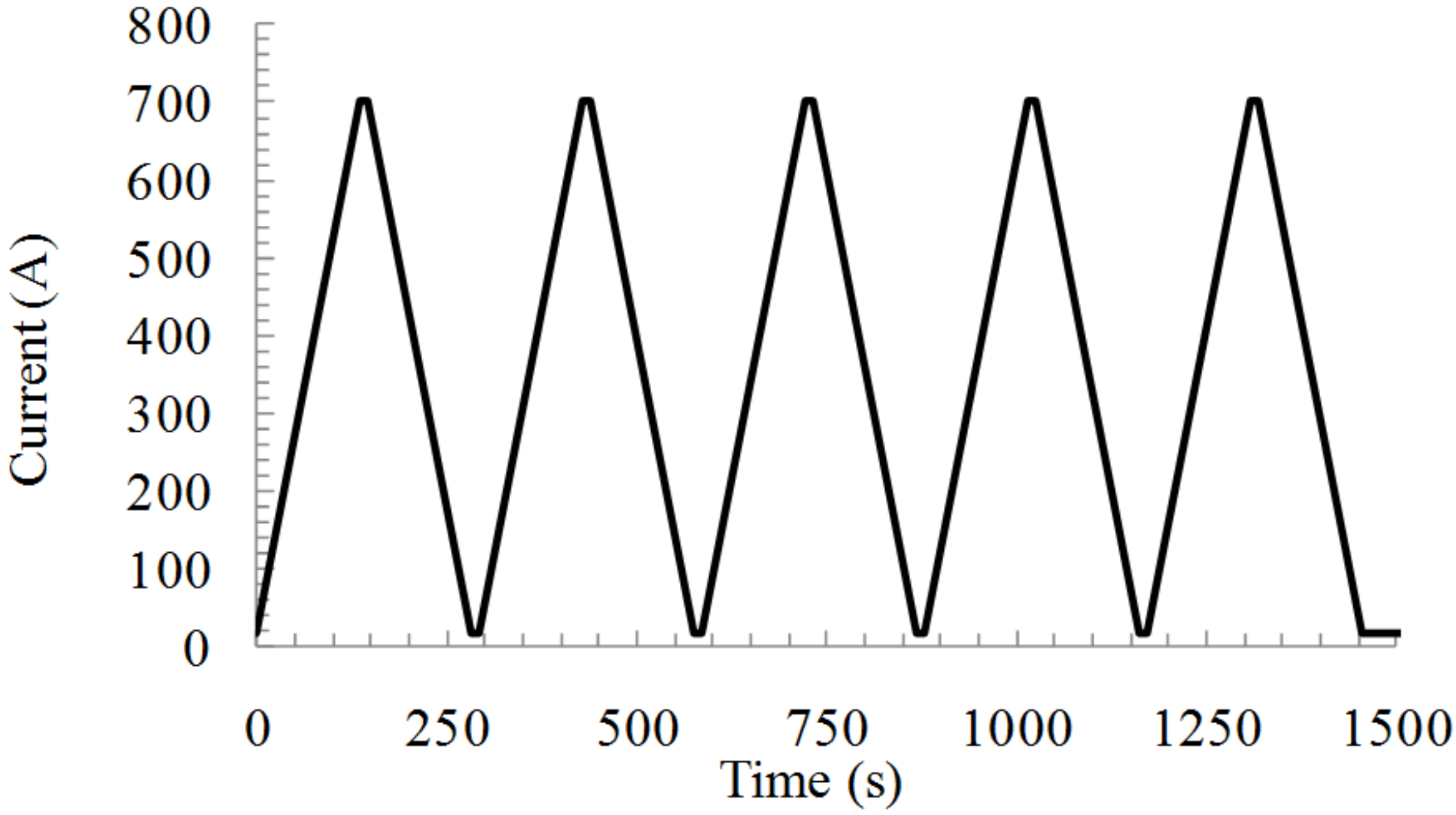

**Fig. 3:** Pre-cycles for LHC warm magnets

When, a transfer line is used at different beam energies, pre-cycles can't always be executed between changes due to time limitations. The degauss is then much more complex and needs advanced techniques. The solution implemented at CERN is to power the magnet at the maximum current for some cycles and then ramp down the current to the nominal value of the beam energy, see Fig. 4.

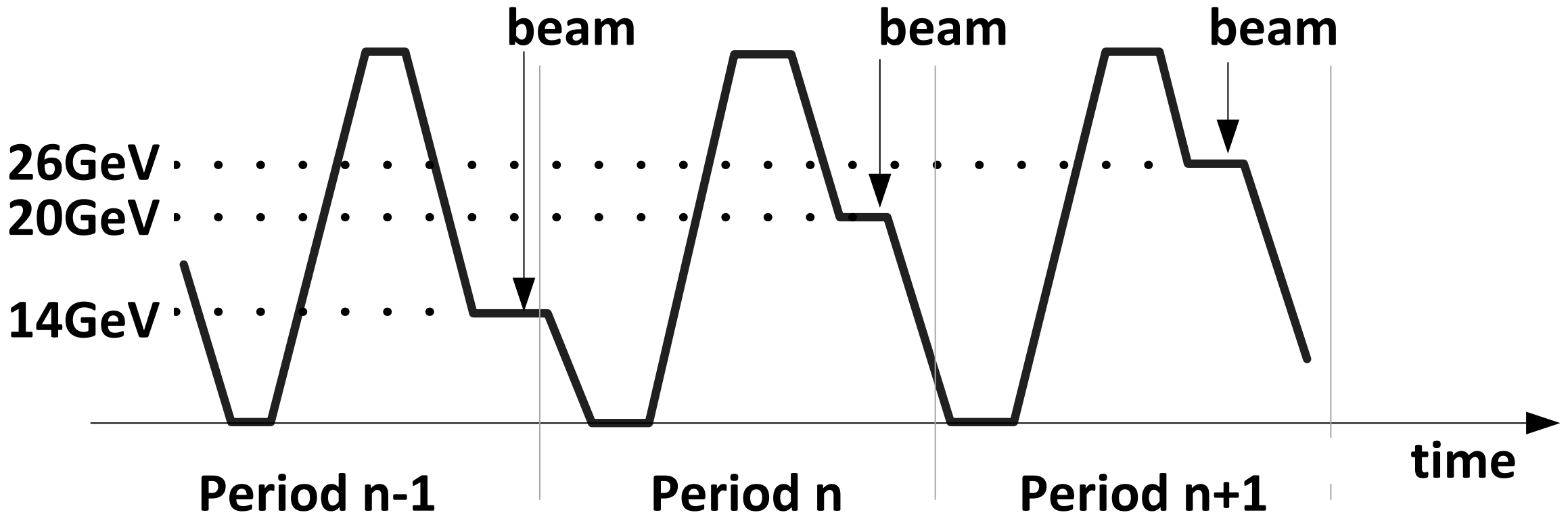

**Fig. 4:** Degauss technique for transfer lines with different beam energies

When the power converters are bipolar in current, the pre-cycle can be done by using an oscillation current shape that allows resetting to zero the magnetic field, see Fig. 5. This is important for corrector magnets for which control is based on the beam position measurement. This kind of pre-cycle is also used for large experimental magnets.

The physical phenomena are even more complex for superconducting magnets. The classical magnetic hysteresis is also present in this case but, in addition, two other phenomena play an important role in magnetic field control. These phenomena are the decay amplitude and snapback, and they are due to the properties of superconducting cables. In this case, a special pre-cycle has to be defined for the LHC main dipoles, see Fig. 6 [4]. In the case of superconducting magnets, pre-cycle is mandatory but is not sufficient to obtain a good control of the magnetic field. The decay and snapback phenomena are so special that a complex model of the magnetic field needs to be built to obtain a good level of prediction [5].

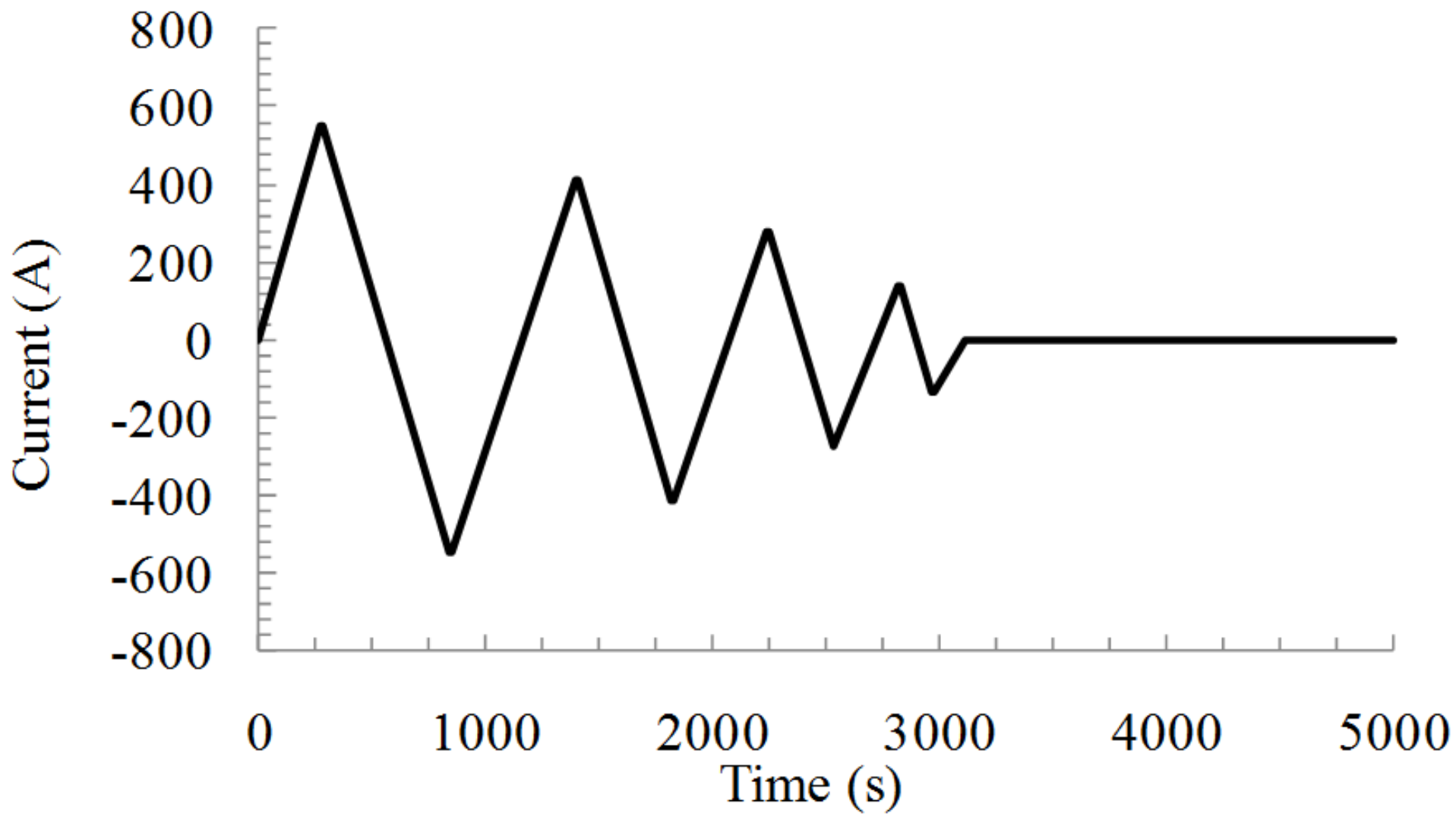

**Fig. 5:** Degauss cycle for corrector magnets

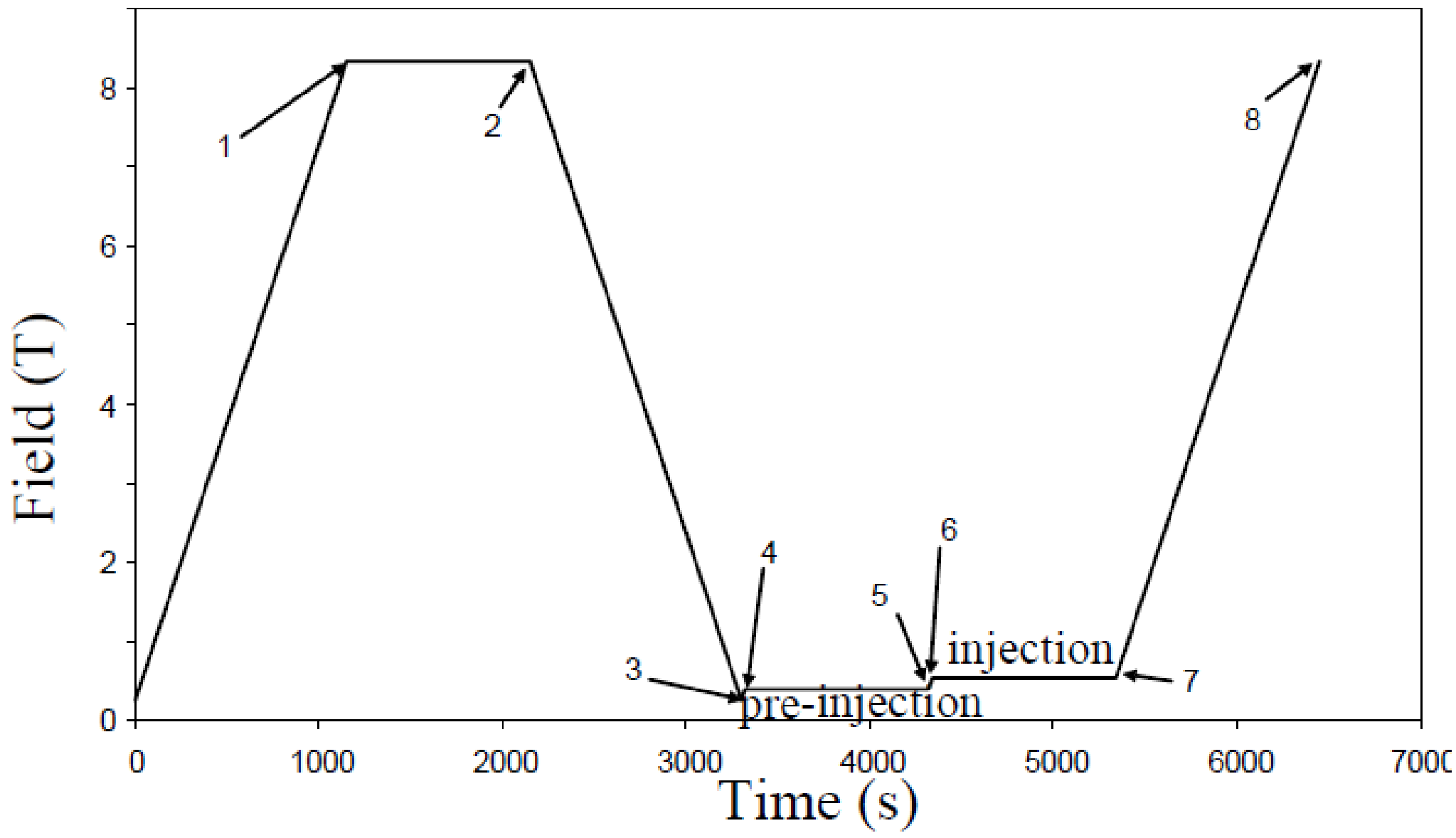

**Fig. 6:** LHC pre-cycle of main dipole magnets

### 2.3 Magnetic field model

LHC operation requires a prediction of the currents for the magnet circuits. These settings are based on a parametric model whose coefficients are obtained from a synthesis of the information available from magnetic field measurements (both room temperature and 1.9 K or 4.2 K). This set of equations, together with the coefficients estimated from measurements, are integrated in a tool at CERN called the Field Description for the LHC (or FiDeL). The aim is to provide the integral transfer function (integral magnetic field versus current) in a form suitable for inversion (current versus integral magnetic field) for each circuit in the LHC [6]. In addition, for the main ring magnets FiDeL provides a prediction of the field errors to be used to set the corrector circuits. FiDeL is implemented in the LHC control system to steer the beam, see Fig. 7. A lot of measurement needs to be done to set up such a tool, but this type of tool is mandatory for the operation of complex and large particle accelerators.

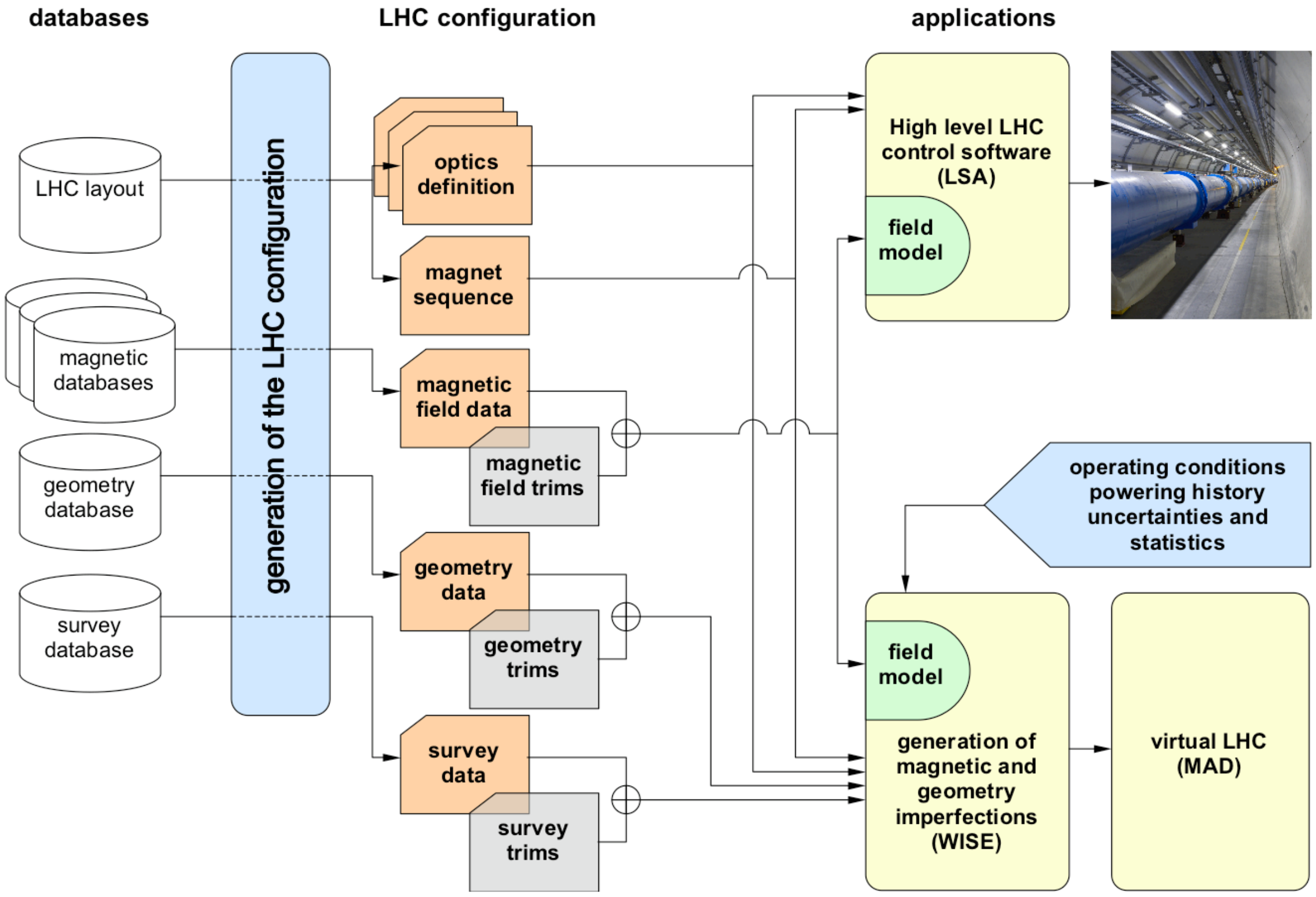

**Fig. 7:** LHC control architecture including FiDeL

The modelling of warm magnets with the same quality of prediction is ongoing at CERN, and it will help the setting up of smaller machines [7].

## 2.4 Magnetic field control

Most of the particle accelerators are current controlled. This is due to the difficulty of measuring in real time the magnetic field of the magnet with the correct level of accuracy. However, since the 1970s this idea has been considered and an additional reference magnet has been added in series with the dipole outside the machine for magnetic transducer instrumentation. This has been the case at CERN for the PS, PSB, and SPS machines. Magnetic field monitoring measurements were made to help with diagnostics during operation.

The PS B-train system is based upon a set of search coils that allow the measurement of the instantaneous field from the coil output voltage. The initial value is provided by a field marker that outputs a trigger pulse when the field reaches a preset level during the pre-injection ramp, thus signalling the start of the integration, see Fig. 8 [8]. In 2008 and for the first time, the PS accelerator was controlled in magnetic field. The magnetic field measurement was connected to the control of the main magnet power system and the regulation was done using this measurement, see Fig. 9 [9]. The global performances are above the expectations due to the improvement of the reproducibility of the machine, which is multi-beam energy.

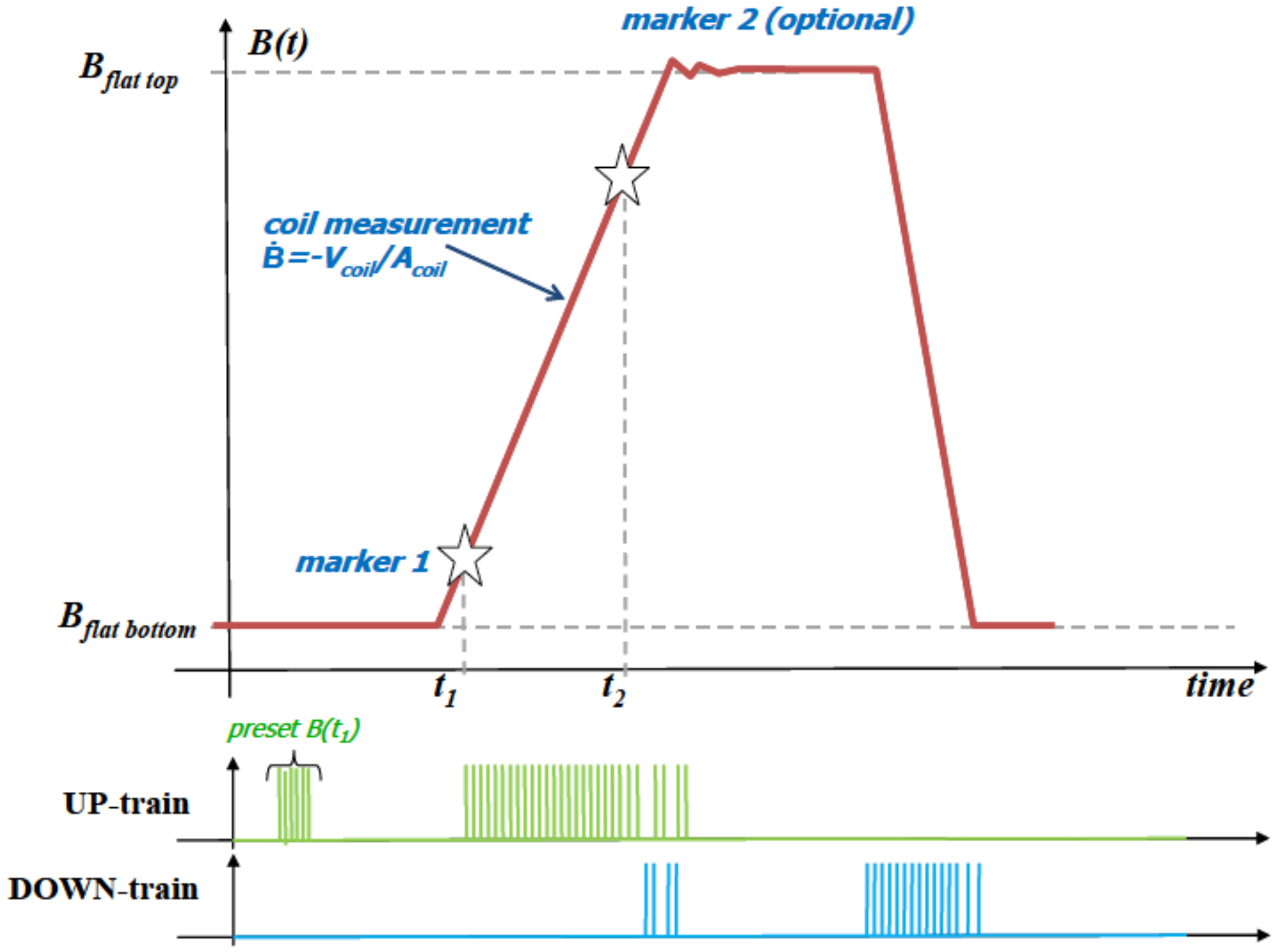

**Fig. 8:** Magnetic cycle with two field markers for the recovery of offset and (optionally) gain errors. A pulse on each one of the two trains represents a ±0.1 G field increment.

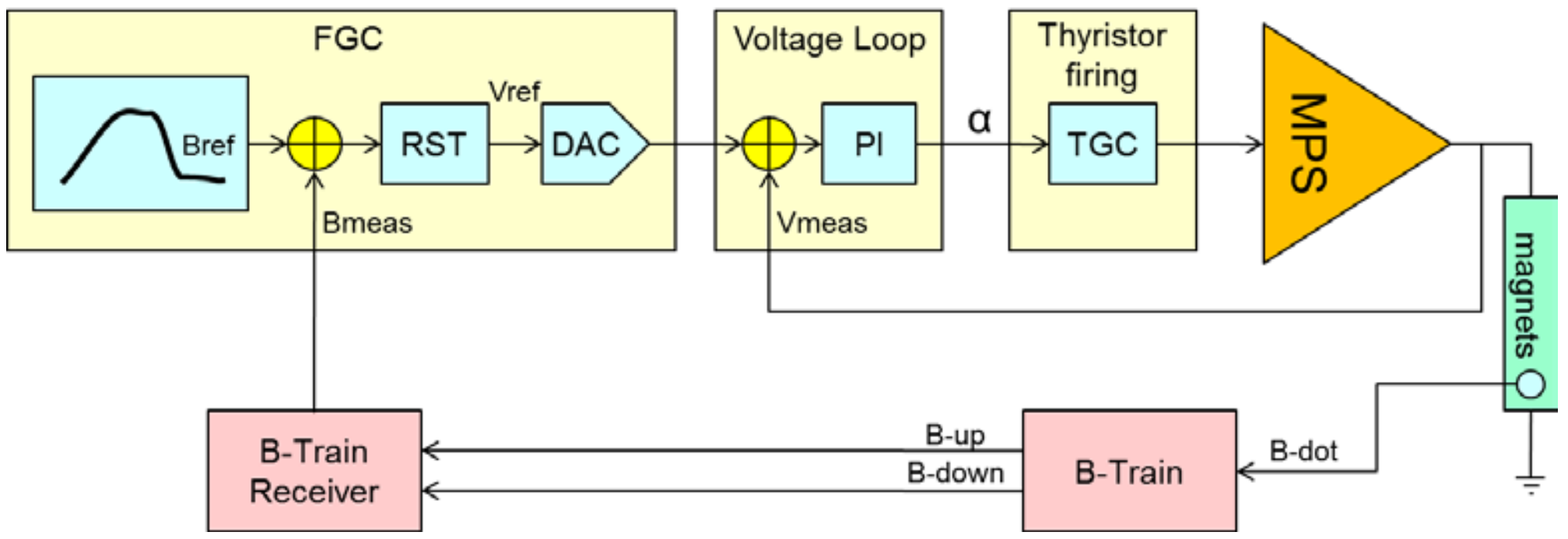

**Fig. 9:** B-field control principle

## 2.5 Orbit feedback system

In some machines, the stability of the beam is critical. It is true for synchrotron light sources such as the LHC, where the presence of two beams, both of high intensity as well as high particle energies, requires excellent control of particle losses inside a superconducting environment, provided by the LHC cleaning and machine protection system [10]. The performance and function of these systems depends critically on the stability of the beam and may eventually limit LHC performance. Environmental and accelerator-inherent sources, as well as the failure of magnets and their power converters, may perturb and reduce beam stability and may consequently lead to an increase in particle losses inside the cryogenic mass. In order to counteract these disturbances, control of the key beam parameters – orbit, tune, energy, coupling, and chromaticity – are an integral part of LHC operation. Since manual correction of these parameters may reach its limit with respect to the required precision and expected timescales, the LHC is the first proton collider that requires feedback control systems for safe and reliable machine operation. The sources of orbit and energy perturbations can be grouped into environmental sources, machine-inherent sources, and machine element failures. The slowest perturbation due to ground motion, tides, and temperature fluctuations in the tunnel can reach 200 μm within 10 hours. These orbit perturbations exceed the required beam stability by one order of magnitude. The LHC cleaning system, imposing one of the tightest constraints on beam stability, requires a beam stability in the range of about 15−25 μm at the location of the collimator jaws. The solution implemented for the LHC is an automated orbit feedback system. It is based on the readings from 1056 beam position monitors (BPMs) that are distributed over the machine; a central global feedback controller calculates new deflection strengths for the more than 1060 orbit corrector magnets (CODs) that correct the orbit and momentum around their references, see Fig. 10. In the LHC the orbit feedback system acts mainly at a low frequency (below 50 Hz), see Fig. 11.

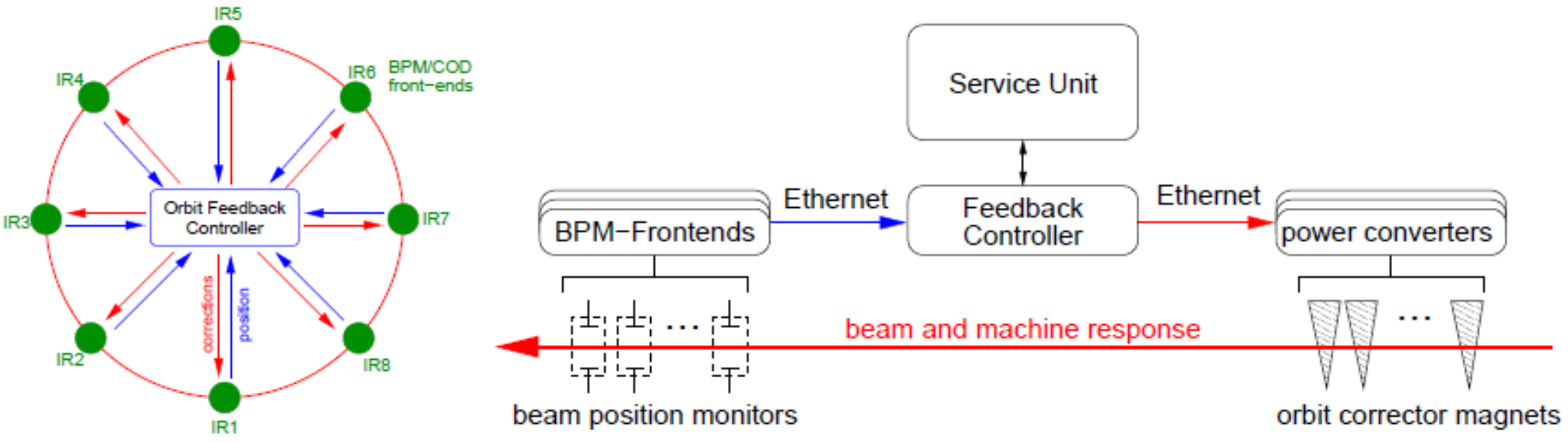

**Fig. 10:** Principle of the LHC orbit feedback system

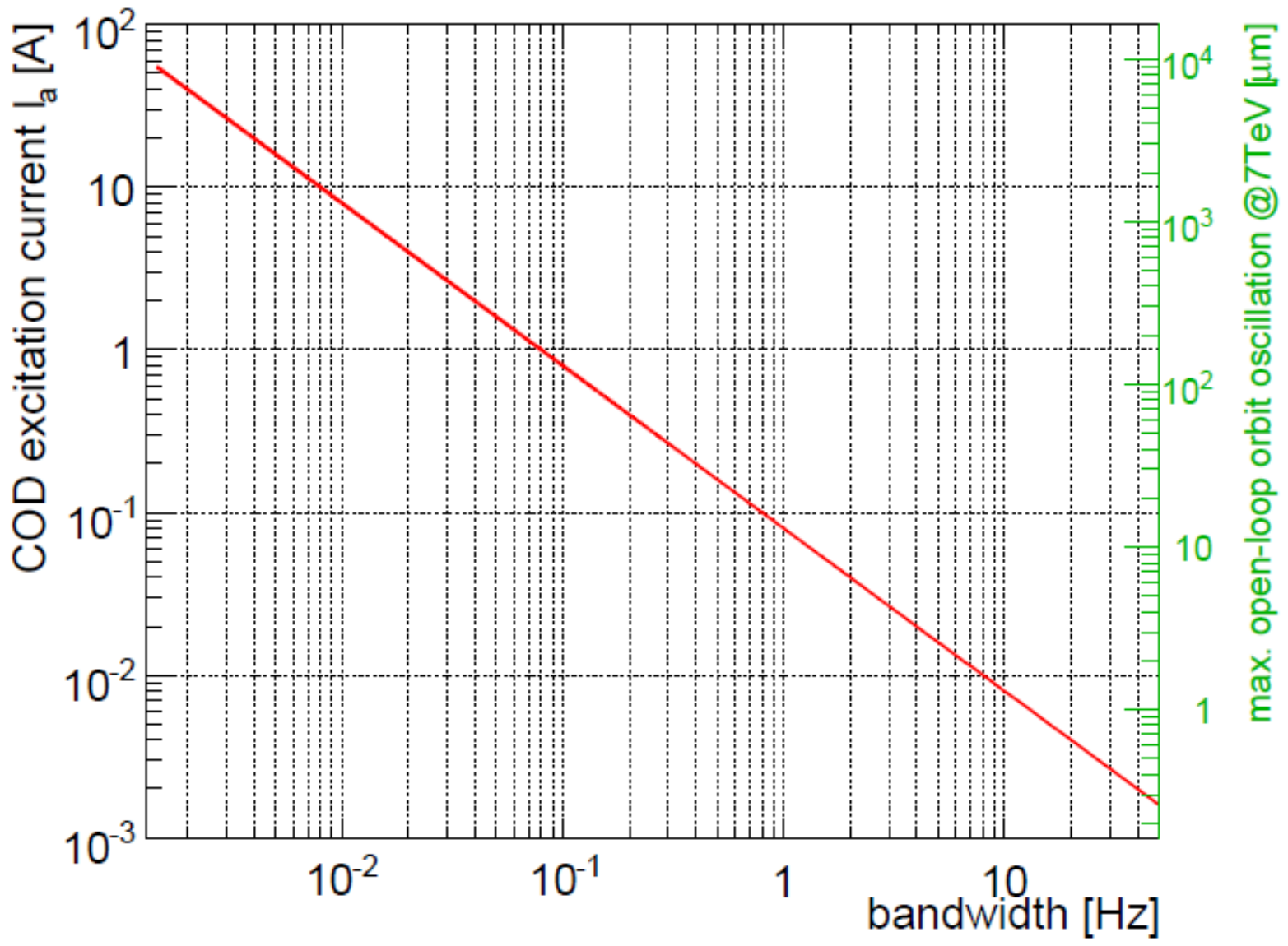

**Fig. 11:** Open-loop bandwidth of the ±60A converter powering the LHC COD magnets. The maximum sinusoidal current amplitude is plotted as a function of bandwidth. The corresponding maximum orbit excursion prediction in the arc driven by the MCB magnet at $\beta$ = 180 m is given on the right-hand scale.

## 3 Power converter control

The control of particle accelerator power converters is one of the most challenging parts and requires special technology. Firstly, it requires high-precision control well above the industrial standard; secondly, all of the power converters for one machine need to be perfectly synchronized (typically with a maximum of 100 μs jitter). However, no standard controls are available on the market to control power converters. For slow applications, many bus protocols may be used (RS422, Ethernet, etc.). For applications that require synchronization, different options could be selected but, unfortunately, no industrial standard control fulfils the requirement of particle accelerators. CERN developed its own controller, called a function generator and controller (FGC), which is a dedicated controller embedded in every power converter [11]. This controller is responsible for function generation (current versus time), current regulation, and power converter state monitoring and control. All of the FGCs are connected to a gateway through a fieldbus, which can be WORLDFIP or Ethernet for the latest generation of FGC. This is needed to link the IP technical network of the machine with the fieldbus segments. Each gateway includes a timing receiver interface, connected to the general machine timing network. This provides the gateways with accurate date/time and millisecond-level machine events. It also generates precise timing pulses to synchronize the WorldFIP interface or to send synchronization timing to the embedded electronics. Each gateway drives a single segment with a maximum of 64 FGCs, see Fig. 12.

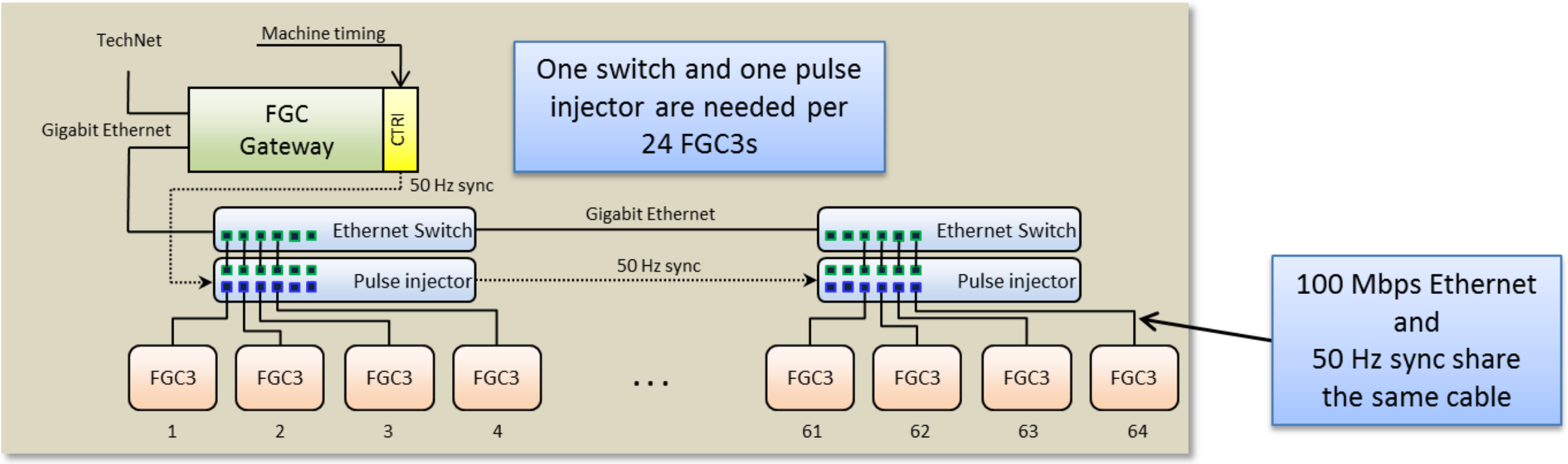

**Fig. 12:** Power converter control architecture at CERN

A key task for each FGC is current regulation. The power part is controlled as a voltage source and the current regulation is done by the FGC. For the LHC, many circuits have very large time constants (many hours) and yet require a very high accuracy, and this can only be achieved with digital regulation. This was the main reason to include a second processor in the design. It operates as a coprocessor for the main microcontroller, performing the real-time tasks of function generation, digital filtering, and current regulation. Thanks to this standardized embedded controller, the software running in all FGCs is the same; only the parameters of the circuits need to be set. The current control algorithm is based on a RST controller, see Fig. 13 [12].

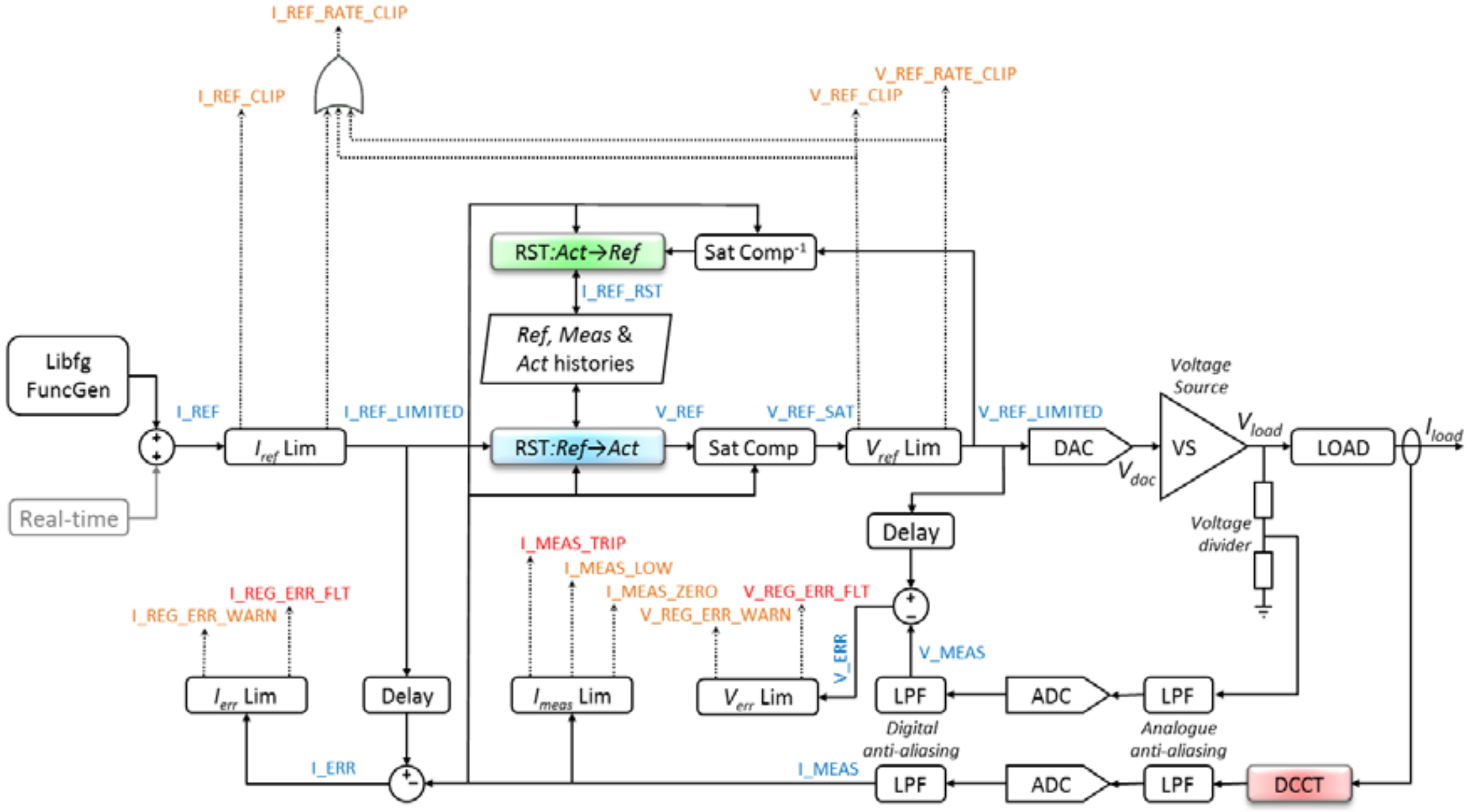

**Fig. 13:** FGC current control architecture

## 4 Control of nested circuits

The control of nested circuits can be challenging. It is particularly true when a change in the current from one power converter generates a perturbation seen by another. In this case, the two power converters are coupled by the load and a special control system has to be designed to stabilize the system. This is the case for the LHC inner triplet magnets, which need three power converters to power four magnets. The control strategy that has been developed is based on a decoupling loop by state feedback control and with three outer independent RST digital current loops, see Fig. 14 [13].

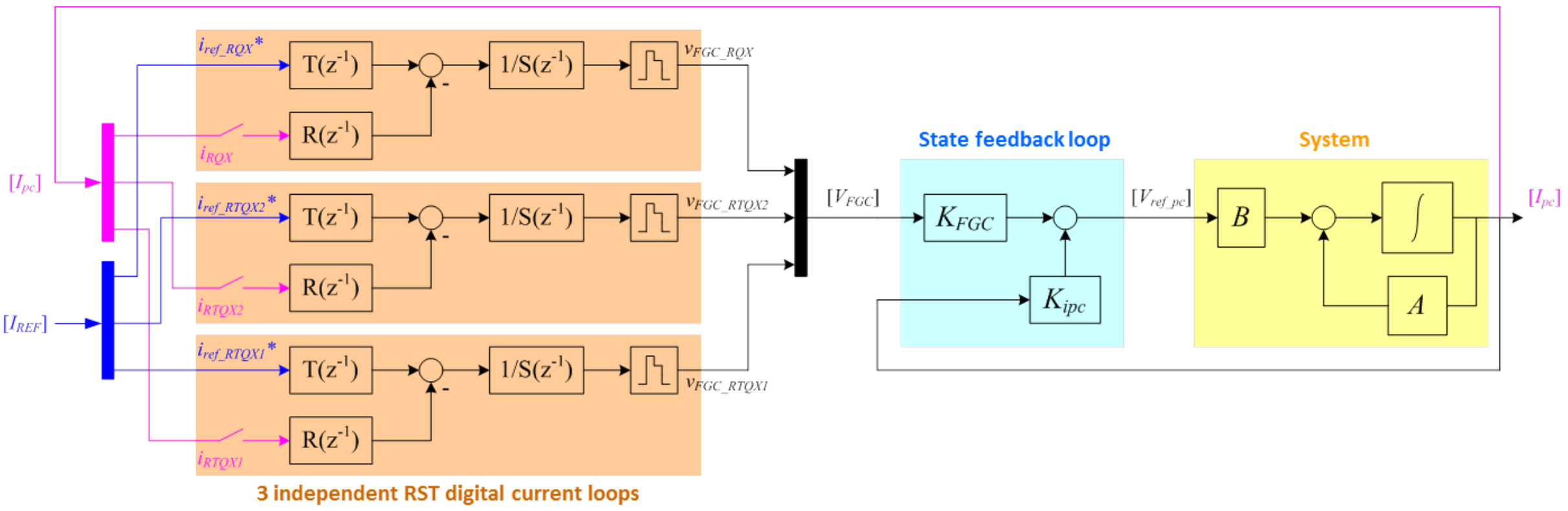

**Fig. 14:** Control strategy for nested circuits

Thank to this control strategy, each power converters can be controlled independently without affecting the stability of the others. The current stability of each power converter stays in the same range as for a single circuit. This decoupling technique eases the operation of nested circuits and it suppresses the need forexpertise while operating such a system.

## 5 Energy saving

Energy consumption is a major concern when operating particle accelerators. Magnets are the main consumer and they must be designed to limit their running costs. At the level of power converters, the efficiency is the main criteria to measure the performance of energy conversion. At the level of the machine, the method of powering the magnets has a huge impact on energy consumption. CERN has developed a new concept to save energy, which is described below.

### 5.1 Pulsing magnets

One of the easiest ways to save energy is to power magnets only when the beam is present inside. This isn't always the case as the easiest way to power a magnet is to keep the current constant all of the time. Moreover, a magnet with a solid yoke can only be powered with a constant current. This solution was widespread as this type of magnet is cheaper than those that are laminated. A classic sight in old facilities are magnets with solid yokes for experimental areas or transfer lines. In these cases, the energy consumption is very high when compared to the real requirement. Energy saving can be evaluated using the ratio of the period of the beam to the repetition time of the beam, see Fig. 15.

A study was done at CERN with the EAST area, where it was demonstrated that electricity consumption could be reduced by 95% by pulsing the magnets [14]. In this case, the beam flat top duration is up to 500 ms and the repetition rate is a few cycles per minute. The power converter types are classical AC/DC converters with enough DC voltage to ramp up the current before the beam's arrival. The return of energy also has to be considered carefully, depending upon the power converter technology.

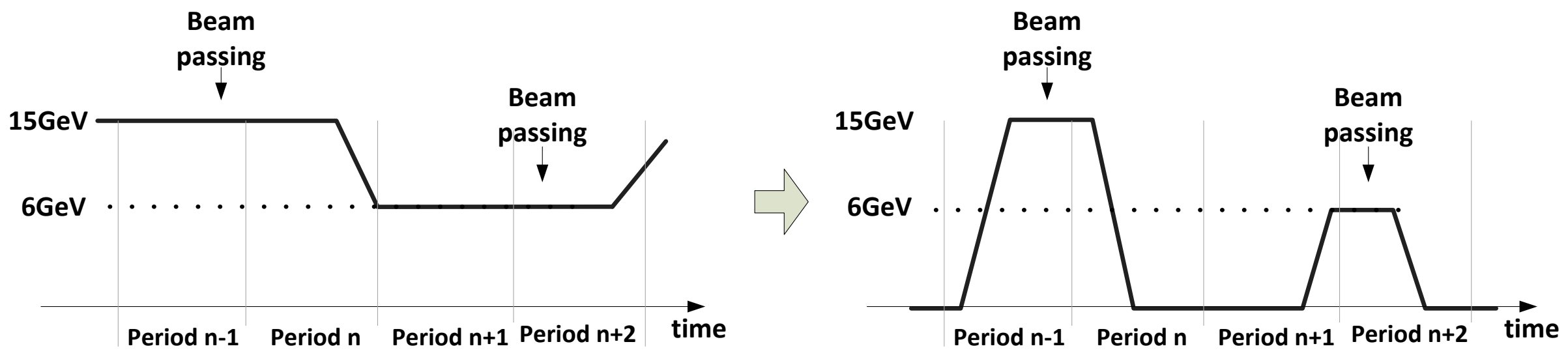

**Fig. 15:** DC operation and pulsed operation

### 5.2 Fast pulsed converters

In the case of linacs, the beam presence in the magnets is in the order of a few milliseconds with a repetition frequency of a few hertz. For example, the beam pulse length is 1.2 ms for Linac4 at CERN with a repetition time of 1.2 s. The beam presence duty cycle is then less than 1%, which leads to an over-consumption if the magnets are powered in DC. A new type of converter was developed at CERN to provide fast trapezoidal current pulses using a high-power insulated-gate bipolar transistor (IGBT) module in its resistive region to regulate the magnet current, see Fig. 16 [15]. Thanks to the pulsed operation, the power converters can be much smaller, and the electricity consumption and cooling capacity requirements are reduced drastically.

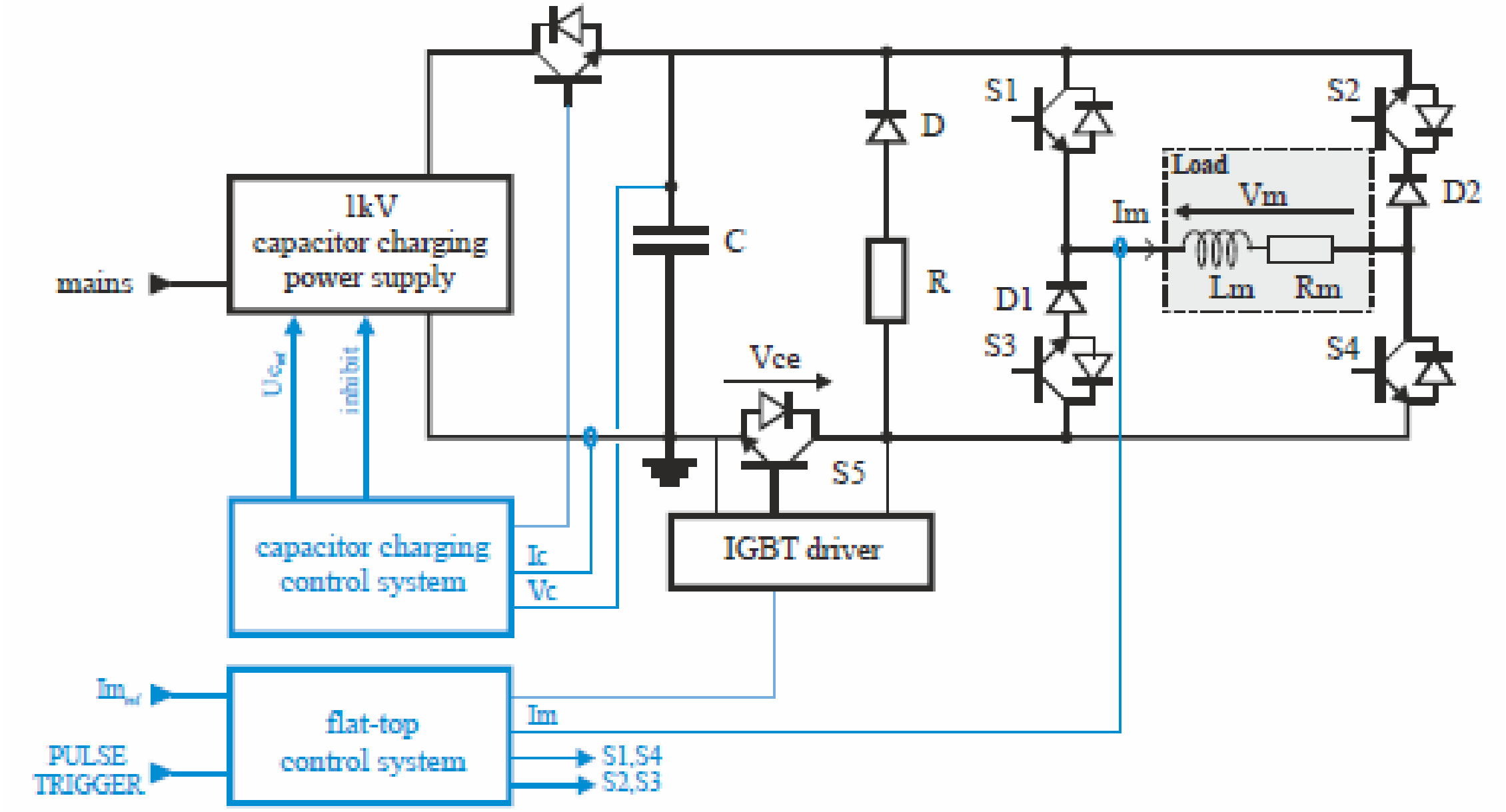

**Fig. 16:** Fast pulsed converter topology

### 5.3 Energy management

Particle accelerators are often characterized by a strong and modulated power demand, which is sometimes incompatible with the power available from the grid. Therefore, large facilities for short-term energy storage have been used, traditionally based on rotating machines. A new concept was introduced at CERN to reduce the power taken from the grid and to limit it to a strict minimum, see Fig. 17. The idea is to store locally energy and to exchange it with the magnets during each cycle [16]. Only the magnets' losses are taken from the grid. Peak power is provided from local energy storage. As particle accelerators are doing thousands of cycles per year, the only possible solution to store energy is within capacitors. Only capacitors can execute millions of charge cycles [17]. The latest optimization was to reduce the power taken from the grid to below the maximum dissipated power in the magnets by anticipating the power demand in each cycle, see Fig. 18.

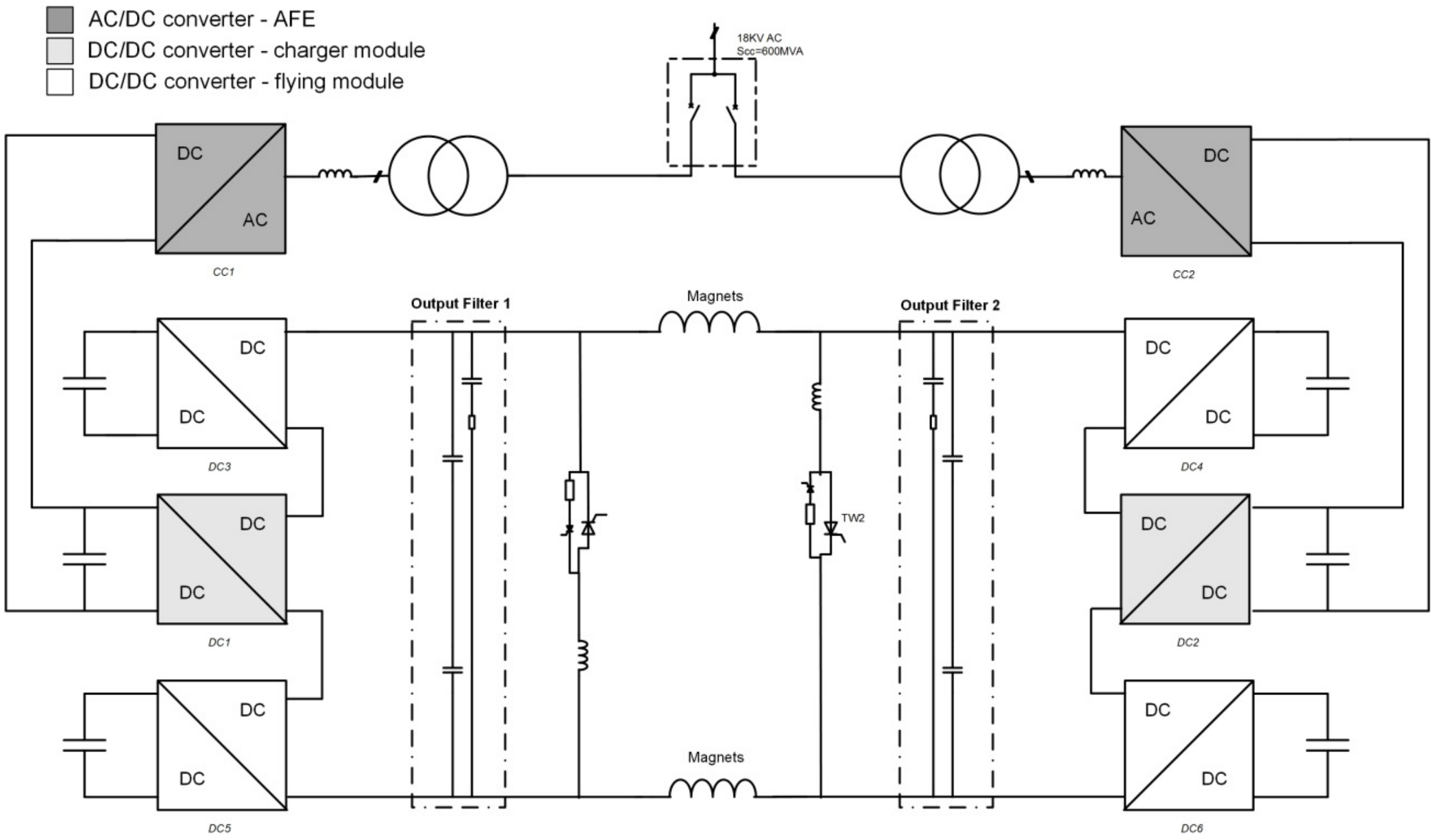

**Fig. 17:** Power converter topology with capacitor energy storage

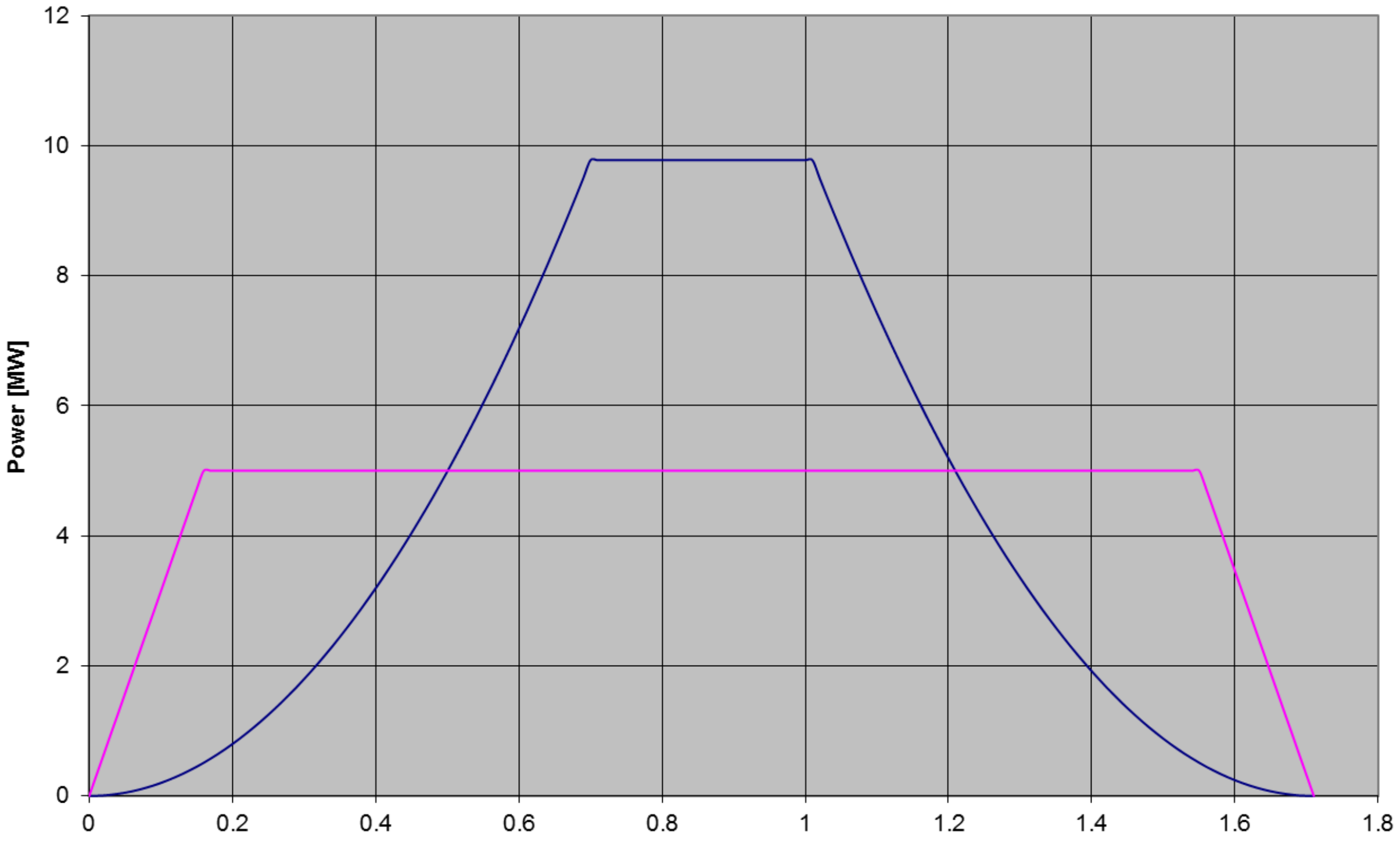

**Fig. 18:** Optimization of power taken from the grid. Blue is the resistive losses of the magnets during a cycle; magenta is the power taken from the grid.

This new concept was introduced at CERN with the power system of the PS accelerator (POPS). It has been in operation since 2011, and has demonstrated its feasibility. However, the stress on the capacitors that are used for energy storage needs to be considered carefully. The discharge cycles, even without any reverse voltage, can degrade the film metallization due to partial discharges. These phenomena need to be taken into account when sizing the capacitor banks [18].

## 6 Summary

This paper presents the latest trends for particle accelerator powering regarding magnetic field uncertainty, power converter control, and energy-saving. These parameters need to be taken into account at the very beginning of any project as they have a major impact on the cost and performance of a machine. They require a system-level approach before the design of power converters, and this can lead to new power converter concepts, like that presented in this paper for saving energy.

## References

[1] Fifty years of the CERN Proton Synchrotron: Vol. 1, edited by S. Gilardoni and D. Manglunki, CERN-2011-004 (CERN, Geneva, 2011), DOI: http://dx.doi.org/10.5170/CERN-2011-004.

[2] H. Thiesen, M. Cerqueira-Bastos, G. Hudson, Q. King, V. Montabonnet, D. Nisbet and S. Page, High precision current control for the LHC main power converters, Proc. IPAC10, Kyoto, Japan, 23-28 May 2010, p. 3260,

http://accelconf.web.cern.ch/AccelConf/IPAC10/papers/wepd070.pdf.

[3] L. Bottura, M. Lamont, E. Todesco, W. Venturini Delsolaro and R. Wolf, Pre-cycles of the LHC magnets during operation, CERN-ATS-2010-174 (CERN, Geneva, 2010).

[4] A. Verweij, N. Sammut and W. Venturini Delsolaro, Pre-cycle selection for the superconducting main magnets of the Large Hadron Collider, Proc. PAC09, Vancouver, 4-8 May 2009, p. 259, http://accelconf.web.cern.ch/AccelConf/PAC2009/papers/mo6pfp054.pdf.

[5] N. Sammut and L. Botura, *Phy. Rev. ST Accel Beams* **12** (2009) 102401.

[6] N. Sammut, L. Botura and J. Micallef, *Phy. Rev. ST Accel Beams* **9** (2006) 012402.

[7] M. Juchno, CERN-Thesis-2009-175.

[8] M. Buzio, P. Galbraith, G. Golluccio, D. Giloteaux, S. Gilardoni, C. Petrone and L. Walckiers, Development of upgraded magnetic instrumentation for CERN real-time reference field measurement systems, Proc. IPAC10, Kyoto, Japan, 23-28 May 2010, p. 310, http://accelconf.web.cern.ch/AccelConf/IPAC10/papers/mopeb016.pdf.

[9] Q. King, S. T. Page and H. Thiesen, Function generation and regulation libraries and their application to the control of the new main power converter (POPS) at the CERN PS, Proc. ICALEPCS2011, Grenoble, France, 10-14 October 2011, p. 886, http://accelconf.web.cern.ch/AccelConf/icalepcs2011/papers/wepmn008.pdf.

[10] R.J. Steinhagen, CERN-Thesis-2007-058.

[11] D. Calcoen, Q. King, P.F. Semanaz, Evolution of the CERN power converter function generator/controller for operation in fast cycling accelerators, Proc. ICALEPCS2011, Grenoble, France, 10-14 October 2011, p. 939,

http://accelconf.web.cern.ch/AccelConf/icalepcs2011/papers/wepmn026.pdf.

[12] F. Bordry and H. Thiesen, RST digital algorithm for controlling the LHC magnet current, CERN-LHC-Project-Report 258, 1998.

[13] F. Bordry, D. Nisbet, H. Thiesen and J. Thomsen, Powering and control strategy for the main quadrupole magnets of the LHC inner triplet system, CERN-ATS-2010-022 (CERN, Geneva, 2011), 2010.

[14] H.J. Burckhart, J-P. Burnet, F. Caspers, V. Doré, L. Gatignon, C. Martel, M. Nonis and D.Tommasini, Some ideas towards energy optimization at CERN, Proc. IPAC13, Shanghai, China, 12-17 May 2013, p. 3400,

http://accelconf.web.cern.ch/AccelConf/IPAC2013/papers/thpfi050.pdf.

[15] J.-M. Cravero and C. De Almeida Martins, A new multiple-stage converter topology for high power and high precision fast pulsed current sources, Proc. EPE09, Barcelona, Spain, 8-10 September 2009 (IEEE, 2009), pp. 1-9.

[16] F. Bordry, J.P. Burnet and F. Voelker, CERN-PS main power converter renovation, how to provide and control the large flow of energy for a rapid cycling machine, Proc. PAC05, Knoxville, United States, 16-20 May 2005, p. 3612,

http://accelconf.web.cern.ch/AccelConf/p05/PAPERS/WPAE063.pdf.

[17] C. Fahrni, A. Rufer, F. Bordry and J.P. Burnet, A novel 60 MW pulsed power system based on capacitive energy storage for particle accelerators, Proc. of EPE07, Aalborg, Denmark, 2-5 September 2007 (IEEE, 2007), p. 1.

[18] F. Boattini and C. Genton, Accelerated lifetime testing of energy storage capacitors used in particle accelerators power converters, EPE15, GENEVA, 8-10 September 2015.

## Bibliography

Proceedings of the CAS-CERN Accelerator School: Power Converters for Particle Accelerators, Montreux, Switzerland, 26-30 March 1990, edited by S. Turner, CERN-1990-007 (CERN, Geneva, 1990), DOI: http://dx/doi.org/10.5170/CERN-1990-007.

Proceedings of the CAS-CERN Accelerator School: Specialised Course on Power Converters, Warrington, United Kingdom, 12-18 May 2004, CERN-2006-010 (CERN, Geneva, 2006), DOI: http://dx.doi.org/10.5170/CERN-2006-010.